\documentclass{PoS}

\title{Dual-frequency VSOP Imaging of a High-redshift Radio Quasar PKS 1402+044}

\ShortTitle{Dual-frequency VSOP Imaging of a High-redshift Radio
Quasar PKS 1402+044}

\author{
        \speaker{Jun Yang}$^{a,f}$, Leonid Gurvits$^b$, Andrei Lobanov$^c$, S\'andor Frey$^{d,e}$ and Xiao-Yu Hong$^{a}$
        \thanks{\textbf{Acknowledgements:}
        This research was partly supported by the Natural Science
        Foundation of China (NSFC10473018 and NSFC10333020).
        Jun Yang and Xiaoyu Hong are grateful to the KNAW--CAS grant.
        S\'andor Frey acknowledges the OTKA T046097 grant. We gratefully
        acknowledge the VSOP Project, which is led by the Institute of
        Space and Astronautical Science (Japan) in cooperation with many
        agencies, institutes and observatories around the world. The
        National Radio Astronomy Observatory is a facility of the National
        Science Foundation operated under cooperative agreement by
        Associated Universities, Inc. This research has made use of NASA's
        Astrophysics Data System, NASA/IPAC Extragalactic Database (NED)
        and the United States Naval Observatory (USNO) Radio Reference
        Frame Image Database (RRFID).} \\
        \llap{$^a$} Shanghai Astronomical Observatory of the Chinese Academy of Sciences \\
                    80 Nandan Road, Shanghai 200030, P.R. China \\
        \llap{$^b$} Joint Institute for VLBI in Europe, P.O. Box 2, 7990 AA Dwingeloo, the Netherlands \\
        \llap{$^c$} Max-Planck-Institut f$\ddot{u}$r Radioastronomie, Auf~dem H$\ddot{u}$gel 69, D-53121 Bonn, Germany \\
        \llap{$^d$} F$\ddot{O}$MI Satellite Geodetic Observatory, P. O. Box 585, H-1592 Budapest, Hungary \\
        \llap{$^e$} MTA Research Group for Physical Geodesy and Geodynamics, Budapest, Hungary \\
        \llap{$^f$} Graduate University of the Chinese Academy of Sciences, Beijing 100049, P.R. China \\
        E-mail: \email{junyang@shao.ac.cn}, \email{lgurvits@jive.nl} \\
                \email{alobanov@mpifr-bonn.mpg.de}, \email{frey@sgo.fomi.hu}, \email{xhong@shao.ac.cn}
        }

\abstract{Based on the VLBI Space Observatory Programme (VSOP)
observations at 1.6 and 5 GHz, we find that the luminous
high-redshift ($z=3.215$) quasar PKS 1402+044 (J1405+0415) has a
pronounced 'core--jet' structure. The jet shows a steeper spectral
index and lower brightness temperature with the increase of the
distance from the core.  The variation of brightness temperature
is basically consistent with the shock-in-jet model. Assuming that
the jet is collimated by the ambient magnetic field, we estimate
the mass of the central object as $\sim10^9M_\odot$. The upper
limit of the jet proper motion of PKS 1402+044 is 0.03
mas~yr$^{-1}$ ($\sim3c$) in the east-west direction.}

\FullConference{8th European VLBI Network Symposium\\
        September 26-29 2006\\
        Toru\'n, Poland}

\begin{document}

\section{Introduction}
High-redshift radio quasars can facilitate a comparison of
structural properties across the redshift space and provide
important inputs into tests of cosmological models, such as the
"apparent angular size--redshift" and "apparent proper
motion--redshift" relations. PKS 1402+044 (J1405+0415) is a
flat-spectrum radio source from the Parkes 2.7-GHz survey (0.58
Jy). Optically it is a 19.6-magnitude stellar object with the
redshift $z=3.215$. The MERLIN observations at 1.6 GHz indicated
that there is a secondary component at a separation of
$0.8^{\prime\prime}$ to the south-west at the position angle of
$-123^{\circ}$ and a faint extended emission at
$3.3^{\prime\prime}$ at the position angle of $-106^\circ$. VLBI
observations \cite{gur92} at 5 GHz found that the main component
consists of a compact core and a resolved jet extending to
$\sim18$ mas to the west. Here we present some results of VSOP
observations at 1.6 and 5 GHz. Throughout the paper, the
cosmological model with $H_0=75$ km s$^{-1}$ Mpc$^{-1}$,
$\Omega_\mathrm{m}=0.3$ and $\Omega_\Lambda=0.7$ is adopted.

\section{Observations and data reduction}
Using the space antenna HALCA and the Very Long Baseline Array
(VLBA), we observed the radio quasar PKS 1402+044 in left circular
polarization for 8 hours at 1.6 GHz on 2001 Jan 21 and for 7 hours
at 5 GHz on 2001 Jan 20. A priori calibrations were done with the
AIPS in a standard way. Fringes were detected on all space--ground
baselines. The useful bandwidth after flaging the side channels
spans for 22.8 MHz. The imaging, self-calibration and model
fitting were done in DIFMAP.

\section{Results and discussion}
\begin{figure*}[h]
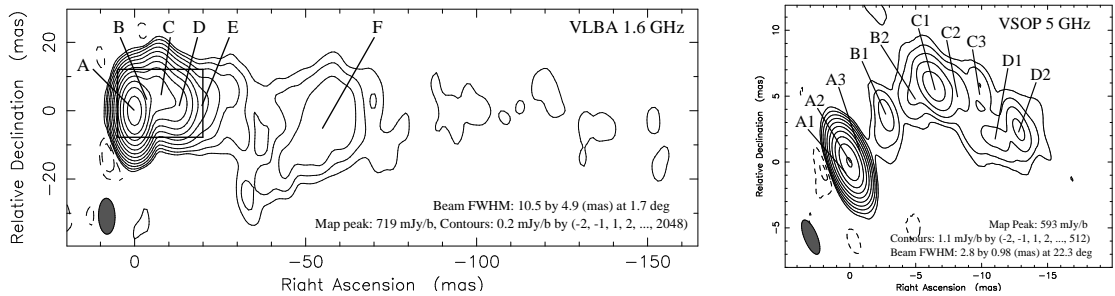

\centering
 \begin{minipage}[c]{0.65\textwidth}
 \includegraphics[trim= 0 80 0 35,  height=3.8cm, clip]{fig1a.eps}
 \end{minipage}
 \begin{minipage}[c]{0.30\textwidth}
 \includegraphics[trim= 0 80  0 38, height=3.8cm,clip]{fig1b.eps}
 \end{minipage}
\caption{VLBI images of PKS 1402+044 (J1405+0415) at 1.6 and 5
GHz. The rectangle in the 1.6-GHz VLBA image shows the area of the
5-GHz VSOP image.} \label{fig1}
\end{figure*}

Figure \ref{fig1} displays a clear core--jet morphology of the
quasar PKS 1402+044. The naturally weighted VLBA image detects the
weak emission extending up to $\sim~150$ mas ($\sim~1$ kpc
projected distance). The jet shows a wide section between 20 and
70 mas (140 -- 500 pc) indicating an expanding jet propagating in
a dense ambient medium. Here we identify the compact core
(component A) and five emission regions (components B -- F). The
naturally weighted 5-GHz VSOP image shows that the inner jet is
resolved into brighter emission regions. With uniform weighting,
the jet is basically resolved out. There is a synchrotron
self-absorbed weak component (A1) appearing at the beginning of
the jet near the brightest component (A2).

Based on the spectral index image between the two frequencies, we
found that the spectral index $\alpha$ ($S_\nu~\propto
\nu^\alpha$) varies from $+0.1$ in the inner nuclear region to
$-1.0$ in the outer jet regions. To further confirm the variation,
we calculated the spectral index of each component using
five-frequency images. The 2.3/8.4-GHz data are from the
RRFID\footnote{USNO Radio Reference Frame Image Database,
http://rorf.usno.navy.mil/RRFID}. The 15-GHz data are from the
VSOP support survey by Gurvits et al. (in preparation). The core
component has a flat spectrum $\alpha=-0.19$, but the jet
components have steeper spectrum: $\alpha_\mathrm{B+C}=-0.55$,
$\alpha_\mathrm{D}=-0.74$. The spectral difference between the
core and jet leads to a decreasing jet to core flux density ratio
with increasing frequency. Furthermore, the difference
demonstrates the explanation of a decreasing jet to core flux
density ratio at a certain observation frequency with increasing
redshift for a large radio quasar sample \cite{fre97}.

If the jet is collimated by the ambient magnetic field
$B_\mathrm{ext}$ ($\sim10^{-5}$ G) of the host galaxy, the mass of
the central object $M_\mathrm{BH}$ can be related to the width of
the jet $r_\mathrm{jet}$ \cite{bes97}: $M_\mathrm{BH}\approx
r_\mathrm{jet}~(B_\mathrm{ext}/B_\mathrm{gr})^{1/2}$ $10^{13}
M_\mathrm{\odot}$, where $B_{gr}$ is the magnetic field at the
Schwarzschild radius. Based on the theoretical assumption
\cite{fie93}, one can expect to have $B_\mathrm{gr}\sim10^4$ G.
Using the measured size 0.3 mas of the component A2, the mass of
the central object is $\sim10^9~M_\mathrm{\odot}$.

For the radio core, the brightness temperature
$T_\mathrm{B}~\approx~4\times10^{12}$~K is close to but somewhat
larger than the inverse Compton limit. Comparing with the limiting
brightness temperature $3\times10^{11}\delta^{5/6}$ K in the
equipartition jet model of Blandford and K\"{o}nigl \cite{bla79},
a lower limit to the Doppler factor $\delta\approx22$ can be
determined. Following the shock-in-jet model of Marscher
\cite{mar90}, we assume that the radio emission is dominated by
adiabatic energy losses. The jet plasma has a power-law energy
distribution, $N(E)dE \propto E^{-s}dE$. The magnetic field varies
as $B \propto d^{-a}$. The Doppler factor is assumed to vary
weakly throughout the jet. There is a simple relation:
$T_\mathrm{B,jet} = T_\mathrm{B,core} (d_\mathrm{jet} /
d_\mathrm{core})^{-\epsilon} $, where $d_\mathrm{jet}$ represents
the measured size of core and jet features and $\epsilon =
[2(2s+1) + 3a(s+1)]/6$. We take $s=2.5$ and $a=1$ corresponding to
the transverse orientation of the magnetic field in the jet
\cite{lob01}. The estimated brightness temperature values are
basically consistent with the observed values.

With another early VLBI observation at 5 GHz in 1986 by Gurvits et
al. \cite{gur92}, we estimated an upper limit of 0.03 mas
yr$^{-1}$ ($\sim3c$) of the proper motion in the EW direction.

\end{document}